# Dynamic Modeling and Statistical Analysis of Event Times

**Edsel A. Peña**

*Abstract.* This review article provides an overview of recent work in the modeling and analysis of recurrent events arising in engineering, reliability, public health, biomedicine and other areas. Recurrent event modeling possesses unique facets making it different and more difficult to handle than single event settings. For instance, the impact of an increasing number of event occurrences needs to be taken into account, the effects of covariates should be considered, potential association among the interevent times within a unit cannot be ignored, and the effects of performed interventions after each event occurrence need to be factored in. A recent general class of models for recurrent events which simultaneously accommodates these aspects is described. Statistical inference methods for this class of models are presented and illustrated through applications to real data sets. Some existing open research problems are described.

## 1. INTRODUCTION

A decade ago in a *Statistical Science* article, Singpurwalla (1995) advocated the development, adoption and exploration of dynamic models in the theory and practice of reliability. He also pinpointed that the use of stochastic processes in the modeling of component and system failure times offers a rich environment to meaningfully capture dynamic operating conditions. In this article, we review recent research developments in dynamic failure-time models, in the context of both stochastic modeling and inference concerning model parameters. Dynamic models are not limited in applicability and relevance to the engineering and reliability areas.

*Edsel A. Peña is Professor, Department of Statistics, University of South Carolina, Columbia, South Carolina 29208, USA e-mail: pena@stat.sc.edu.*

They are also relevant in other fields such as public health, biomedicine, economics, sociology and politics. This is because in many studies in these varied areas, it is oftentimes of interest to monitor the occurrences of an event. Such an event could be the malfunctioning of a mechanical or electronic system, encountering a bug in computer software, the outbreak of a disease, occurrence of a migraine, reoccurrence of a tumor, a drop of 200 points in the Dow Jones Industrial Average during a trading day, commission of a criminal act in a city, serious disagreements between a married couple, or the replacement of a cabinet minister/secretary in an administration. These events recur and so it is of interest to describe their recurrence behavior through a stochastic model. If such a model has excellent predictive ability for event occurrences, then if event occurrences lead to drastic and/or negative consequences, preventive interventions may be attempted to minimize damages, whereas if they lead to beneficial and/or positive outcomes, certain actions may be performed to hasten event occurrences.

It is because of performed interventions after event occurrences that dynamic models become especially pertinent, since through such interventions, or sometimes noninterventions, the stochastic structure governing future event occurrences is altered. This change in the mechanism governing the event occurrences could also be due to the induced change in the structure function in a reliability setting governing the system of interest arising from an event occurrence (cf. Hollander and Peña, 1995; Kvam and Peña, 2005; Peña and Slate, 2005). Furthermore, since several events may occur within a unit, it is also important to consider the association among the interevent times which may arise because of unobserved random effects or frailties for the unit. In addition, there is a need to take into account the potential impact of environmental and other relevant covariates, possibly including the accumulated number of event occurrences, which could affect future event occurrences.

In this review article we describe a flexible and general class of dynamic stochastic models for event occurrences. This class of models was proposed by Peña and Hollander (2004). We also discuss inference methods for this class of models as developed in Peña, Strawderman and Hollander (2001), Peña, Slate and González (2007) and Kvam and Peña (2005). We demonstrate their applications to real data sets and indicate some open research problems.

## 2. BACKGROUND AND NOTATION

Rapid and general progress in event time modeling and inference has benefited immensely from the adoption of the framework of counting processes, martingales and stochastic integration as introduced in Aalen's (1978) pioneering work. The present review article adopts this mathematical framework. Excellent references for this framework are the monograph of Gill (1980), and the books by Fleming and Harrington (1991) and Andersen, Borgan, Gill and Keiding (1993). We introduce in this section some notation and very minimal background in order to help the reader in the sequel, but advise the reader to consult the aforementioned references to gain an in-depth knowledge of this framework.

We denote by $(\Omega, \mathcal{F}, \mathbf{P})$ the basic probability space on which all random entities are defined. Since interest will be on times between event occurrences or on the times of event occurrences, nonnegative-valued random variables will be of concern. For a random variable $T$ with range $\Re_+ = [0, \infty)$, $F(t) = F_T(t) = \mathbf{P}\{T \leq t\}$ and $S(t) = S_T(t) = 1 - F(t) = \mathbf{P}\{T > t\}$ will denote its distribution and survivor (also called reliability) functions, respectively. The indicator function of the event $A$ will be denoted by $I\{A\}$. The cumulative hazard function of $T$ is defined according to

$$\Lambda(t) = \Lambda_T(t) = I\{t \geq 0\} \int_0^t \frac{F(dw)}{S(w-)}$$

with the convention that $S(w-) = \lim_{t \uparrow w} S(t) = \mathbf{P}\{T \geq w\}$ and $\int_a^b \cdots = \int_{(a,b]} \cdots$. For a nondecreasing function $G: \Re_+ \to \Re_+$ with $G(0) = 0$, its product-integral over $(0, t]$ is defined via (cf. Gill and Johansen, 1990; Andersen et al., 1993) $\prod_{w=0}^t [1 - G(dw)] = \lim_{M \to \infty} \prod_{i=1}^M [1 - (G(t_i) - G(t_{i-1}))]$, where as $M \to \infty$, the partition $0 = t_0 < t_1 < \cdots < t_M = t_0$ satisfies $\max_{1 \leq i \leq M} |t_i - t_{i-1}| \to 0$. The survivor function in terms of the cumulative hazard function becomes

$$(2.1) \quad S(t) = I\{t < 0\} + I\{t \geq 0\} \prod_{w=0}^t [1 - \Lambda(dw)].$$

For a discrete random variable $T$ with jump points $\{t_j\}$, the hazard $\lambda_j$ at $t_j$ is the conditional probability that $T = t_j$, given $T \geq t_j$, so $\Lambda(t) = \sum_{\{j : t_j \leq t\}} \lambda_j$. If $T$ is an absolutely continuous random variable with density function $f(t)$, its hazard rate function is $\lambda(t) = f(t)/S(t)$, so $\Lambda(t) = \int_0^t \lambda(w) \, dw = -\log S(t)$. The product-integral representation of $S(t)$ according to whether $T$ is purely discrete or purely continuous is

$$\begin{aligned}
S(t) &= \prod_{w=0}^t [1 - \Lambda(dw)] \\
(2.2) \\
&= \begin{cases} \prod_{t_j \leq t} [1 - \lambda_j], & \text{if } T \text{ is discrete,} \\ \exp\left\{-\int_0^t \lambda(w) \, dw\right\}, & \text{if } T \text{ is continuous.} \end{cases}
\end{aligned}$$

A benefit of using hazards or hazard rate functions in modeling is that they provide qualitative aspects of the event occurrence process as time progresses. For instance, if the hazard rate function is increasing (decreasing) then this indicates that the event occurrences are increasing (decreasing) as time increases, and thus we have the notion of increasing (decreasing) failure rate [IFR (DFR)] distributions. For many years, it was the focus of theoretical reliability research to deal with classes of distributions such as IFR, DFR, increasing (decreasing) failure rate on average [IFRA (DFRA)], and so on, specifically with regard to their closure properties under certain reliability operations (cf. Barlow and Proschan, 1981).

In monitoring an experimental unit for the occurrence of a recurrent event, it is convenient and advantageous to utilize a counting process $\{N(s), s \geq 0\}$, where $N(s)$ denotes the number of times that the event has occurred over the interval $[0, s]$. The paths of this stochastic process are step-functions with $N(0) = 0$ and with jumps of unity. If we represent the calendar times of event occurrences by $S_1 < S_2 < S_3 < \cdots$ with the convention that $S_0 = 0$, then $N(s) = \sum_{j=1}^{\infty} I\{S_j \leq s\}$. The interevent times are denoted by $T_j = S_j - S_{j-1}, j = 1, 2, \ldots$. In specifying the stochastic characteristics of the event occurrence process, one either specifies all the finite-dimensional distributions of the process $\{N(s)\}$, or specifies the joint distributions of the $S_j$'s or the $T_j$'s. For example, a common specification for event occurrences is the assumption of a homogeneous Poisson process (HPP) where the interevent times $T_j$ are independent and identically distributed exponential random variables with scale $\beta$. This is equivalent to specifying that, for any $s_1 < s_2 < \cdots < s_K$, the random vectors $(N(s_1), N(s_2) - N(s_1), \ldots, N(s_K) - N(s_{K-1}))$ have independent components with $N(s_j) - N(s_{j-1})$ having a Poisson distribution with parameter $\beta(s_j - s_{j-1})$. From this specification, the finite-dimensional distributions of $(N(s_1), N(s_2), \ldots, N(s_K))$ can be obtained.

An important concept in dynamic modeling is that of a history or a filtration, a family of $\sigma$-fields $\mathbf{F} = \{\mathcal{F}_s : s \geq 0\}$ where $\mathcal{F}_s$ is a sub-$\sigma$-field of $\mathcal{F}$ with $\mathcal{F}_{s_1} \subseteq \mathcal{F}_{s_2}$ for every $s_1 < s_2$ and with $\mathcal{F}_s = \bigcap_{h \downarrow 0} \mathcal{F}_{s+h}$, a right-continuity property. One interprets $\mathcal{F}_s$ as all information that accrued over $[0, s]$. When augmented in $(\Omega, \mathcal{F}, \mathbf{P})$, we obtain the filtered probability space $(\Omega, \mathcal{F}, \mathbf{F}, \mathbf{P})$. It is with respect to such a filtered probability space that a process $\{X(s) : s \geq 0\}$ is said to be adapted [$X(s)$ is measurable with respect to $\mathcal{F}_s, \forall s \geq 0$]; is said to be a (sub)martingale [adapted, $E|X(s)| < \infty$, and, $\forall s, t \geq 0, E\{X(s+t)|\mathcal{F}_s\} (\geq) = X(s)$ almost surely]. Doob–Meyer's decomposition guarantees that for a submartingale $\mathbf{Y} = \{Y(s) : s \geq 0\}$ there is a unique increasing predictable (measurable with respect to the $\sigma$-field generated by adapted processes with left-continuous paths) process $\mathbf{A} = \{A(s) : s \geq 0\}$, called the compensator, with $A(0) = 0$ such that $\mathbf{M} = \{M(s) = Y(s) - A(s) : s \geq 0\}$ is a martingale. Via Jensen's inequality, then for a square-integrable martingale $\mathbf{X} = \{X(s) : s \geq 0\}$, there is a unique compensator process $\mathbf{A} = \{A(s) : s \geq 0\}$ such that $\mathbf{X}^2 - \mathbf{A}$ is a martingale. This process $\mathbf{A}$, denoted by $\langle \mathbf{X} \rangle = \{\langle X \rangle(s) : s \geq 0\}$, is called the predictable quadratic variation (PQV) process of $\mathbf{X}$. A useful heuristic, though somewhat imprecise, way of presenting the main properties of martingales and PQV's is through the following differential forms. For a martingale $\mathbf{M}$, $E\{dM(s)|\mathcal{F}_{s-}\} = 0, \forall s \geq 0$; whereas for the PQV $\langle \mathbf{M} \rangle$, $E\{dM^2(s)|\mathcal{F}_{s-}\} = \mathrm{Var}\{dM(s)|\mathcal{F}_{s-}\} = d\langle M \rangle(s), \forall s \geq 0$.

For the HPP $\mathbf{N} = \{N(s) : s \geq 0\}$ with rate $\beta$ and with $\mathbf{F} = \{\mathcal{F}_s = \sigma\{N(w), w \leq s\} : s \geq 0\}$, $\mathbf{N}$ is a submartingale since its paths are nondecreasing. Its compensator process is $\mathbf{A} = \{A(s) = \beta s : s \geq 0\}$, so that $\mathbf{M} = \{M(s) = N(s) - \beta s : s \geq 0\}$ is a martingale. Furthermore, since $N(s)$ is Poisson-distributed with rate $\beta s$, so that $\{M^2(s) - A(s) = (N(s) - \beta s)^2 - \beta s : s \geq 0\}$ is a martingale, the PQV process of $\mathbf{M}$ is also $\mathbf{A}$. Through the heuristic forms, we have $E\{dN(s)|\mathcal{F}_{s-}\} = dA(s), s \geq 0$. Since $dN(s) \in \{0, 1\}$, then we obtain the probabilistic expression $\mathbf{P}\{dN(s) = 1|\mathcal{F}_{s-}\} = dA(s), s \geq 0$. Analogously, $\mathrm{Var}\{dN(s)|\mathcal{F}_{s-}\} = E\{[dN(s) - dA(s)]^2|\mathcal{F}_{s-}\} = dA(s), s \geq 0$. These formulas hold for a general counting process $\{N(s) : s \geq 0\}$ with compensator process $\{A(s) : s \geq 0\}$. In essence, conditionally on $\mathcal{F}_{s-}$, $dN(s)$ has a Bernoulli distribution with success probability $dA(s)$. Over the interval $[0, s]$, following Jacod, the likelihood function can be written in product-integral form as

$$L(s) = \prod_{w=0}^{s} \{dA(w)\}^{dN(w)} \{1 - dA(w)\}^{1-dN(w)}$$
$$(2.3)$$
$$= \left\{ \prod_{w=0}^{s} [dA(w)]^{dN(w)} \right\} \exp\{-A(s)\},$$

with the last equality holding when $A(\cdot)$ has continuous paths.

Stochastic integrals play a crucial role in this stochastic process framework for event time modeling. For a square-integrable martingale $\mathbf{X} = \{X(s) : s \geq 0\}$ with PQV process $\langle \mathbf{X} \rangle = \{\langle X \rangle(s) : s \geq 0\}$, and for a bounded predictable process $Y = \{Y(s) : s \geq 0\}$, the stochastic integral of $Y$ with respect to $X$, denoted by $\int Y \, dX = \{\int_0^s Y(w) \, dX(w) : s \geq 0\}$, is well defined. It is also a square-integrable martingale with PQV process $\langle \int Y \, dX \rangle = \{\int_0^s Y^2(w) \, d\langle X \rangle(w) : s \geq 0\}$. When $X$ is associated with a counting process $N$, that is, $X = N - A$, the paths of the stochastic integral $\int Y \, dX$ can be taken as pathwise Lebesgue–Stieltjes integrals.

Martingale theory also plays a major role in obtaining asymptotic properties of estimators as first demonstrated in Aalen (1978), Gill (1980) and Andersen and Gill (1982). The main tools used in asymptotic analysis are Lenglart's inequality (cf. Lenglart, 1977; Fleming and Harrington, 1991; Andersen et al., 1993) which is used in proving consistency, and Rebolledo's (1980) martingale central limit theorem (MCLT) (cf. Fleming and Harrington, 1991; Andersen et al., 1993) which is used for obtaining weak convergence results. We refer the reader to Fleming and Harrington (1991) and Andersen et al. (1993) for the in-depth theory and applications of these modern tools in failure-time analysis.

## 3. CLASS OF DYNAMIC MODELS

Let us now consider a unit being monitored over time for the occurrence of a recurrent event. The monitoring period could be a fixed interval or it could be a random interval, and for our purposes we denote this period by $[0, \tau]$, where $\tau$ is assumed to have some distribution $G$, which may be degenerate. With a slight notational change from Section 2 we denote by $N^\dagger(s)$ the number of events that have occurred on or before time $s$, and by $Y^\dagger(s)$ an indicator process which equals 1 when the unit is still under observation at time $s$, 0 otherwise. With $S_0 = 0$ and $S_j, j = 1, 2, \ldots$, denoting the successive calendar times of event occurrences, and with $T_j = S_j - S_{j-1}, j = 1, 2, \ldots$, being the interevent or gap times, observe that

$$\begin{aligned}N^\dagger(s) &= \sum_{j=1}^\infty I\{S_j \leq \min(s, \tau)\} \quad \text{and} \\ Y^\dagger(s) &= I\{\tau \geq s\}.\end{aligned} \quad (3.4)$$

Associated with the unit is a, possibly time-dependent, $1 \times q$ covariate vector $\mathbf{X} = \{\mathbf{X}(s) : s \geq 0\}$. In reliability engineering studies, the components of this covariate vector could be related to environmental or operating condition characteristics; in biomedical studies, they could be blood pressure, treatment assigned, initial tumor size, and so on; in a sociological study of marital disharmony, they could be length of marriage, family income level, number of children residing with the couple, ages of children, and so on. Usually, after each event occurrence, some form of intervention is applied or performed, such as replacing or repairing failed components in a reliability system, or reducing or increasing physical activity after a heart attack in a medical setting. These interventions will typically impact the next occurrence of the event. There is furthermore recognition that for the unit the interevent times are associated or correlated, possibly because of unobserved random effects or so-called frailties. A pictorial representation of these aspects is given in Figure 1. Observe that because of the finiteness of the monitoring period, which leads to a sum-quota accrual scheme, there will always be a right-censored interevent time. The observed number of event occurrences over $[0, \tau]$, $K = N^\dagger(\tau)$, is also informative about the stochastic mechanism governing event occurrences. In fact, since $K = \max\{k : \sum_{j=1}^k T_j \leq \tau\}$, then, conditionally on $(K, \tau)$, the vector $(T_1, T_2, \ldots, T_K)$ has dependent components, even if at the outset $T_1, T_2, \ldots$ are independent.

Recognizing these different aspects in recurrent event settings, Peña and Hollander (2004) proposed a general class of models that simultaneously incorporates all of these aspects. To describe this class of models, we suppose that there is a filtration $\mathbf{F} = \{\mathcal{F}_s : s \geq 0\}$ such that for each $s \geq 0$, $\sigma\{(N^\dagger(v), Y^\dagger(v+), X(v+), \mathcal{E}(v+)) : v \leq s\} \subseteq \mathcal{F}_s$. We also assume that there exists an unobservable positive random variable $Z$, called a frailty, which induces the correlation among the inter-event times. The class of models of Peña and Hollander (2004) can now be described in differential form via

$$\begin{aligned}(3.5) \quad &\mathbf{P}\{dN(s) = 1 | \mathcal{F}_{s-}, Z\} \\ &= Z Y^\dagger(s) \lambda_0[\mathcal{E}(s)] \\ &\quad \cdot \rho[N^\dagger(s-); \alpha] \psi(\mathbf{X}(s)\beta) \, ds,\end{aligned}$$

where $\lambda_0(\cdot)$ is a baseline hazard rate function; $\rho(\cdot; \cdot)$ is a nonnegative function with $\rho(0; \cdot) = 1$ and with $\alpha$ being some parameter; $\psi(\cdot)$ is a nonnegative link function with $\beta$ a $q \times 1$ regression parameter vector; and $Z$ is a frailty variable. The at-risk process $Y^\dagger(s)$ indicates that the conditional probability of an event occurring becomes zero whenever the unit is not under observation. Possible choices of the $\rho(\cdot; \cdot)$ and $\psi(\cdot)$ functions are $\rho(k; \alpha) = \alpha^k$ and $\psi(w) = \exp(w)$, respectively. For the geometric choice of $\rho(\cdot; \cdot)$, if $\alpha > 1$ the effect of accumulating event occurrences is to accelerate event occurrences, whereas if $\alpha < 1$ the event occurrences decelerate, the latter situation appropriate in software debugging. The process $\mathcal{E}(\cdot)$ appearing as argument in the baseline hazard rate function, called the effective age process, is predictable, observable, nonnegative and pathwise almost surely differentiable with derivative $\mathcal{E}'(\cdot)$. This effective age process models the impact of performed interventions after each event occurrence. A pictorial depiction of this effective age process is in Figure 2. In this plot, after the first event, the performed intervention has the effect of reverting the unit to an effective age equal to the age just before the event occurrence (called a minimal repair or intervention), while after the second event, the performed intervention has the effect of reverting the effective age to that of a new unit (hence this is called a perfect intervention or repair). After the third event, the intervention is neither minimal nor perfect and it has the effect of restarting the effective age at a value between zero and that just before the event occurred; while for the fourth event, the intervention is detrimental in that the restarting effective age exceeds that just before the event occurred.

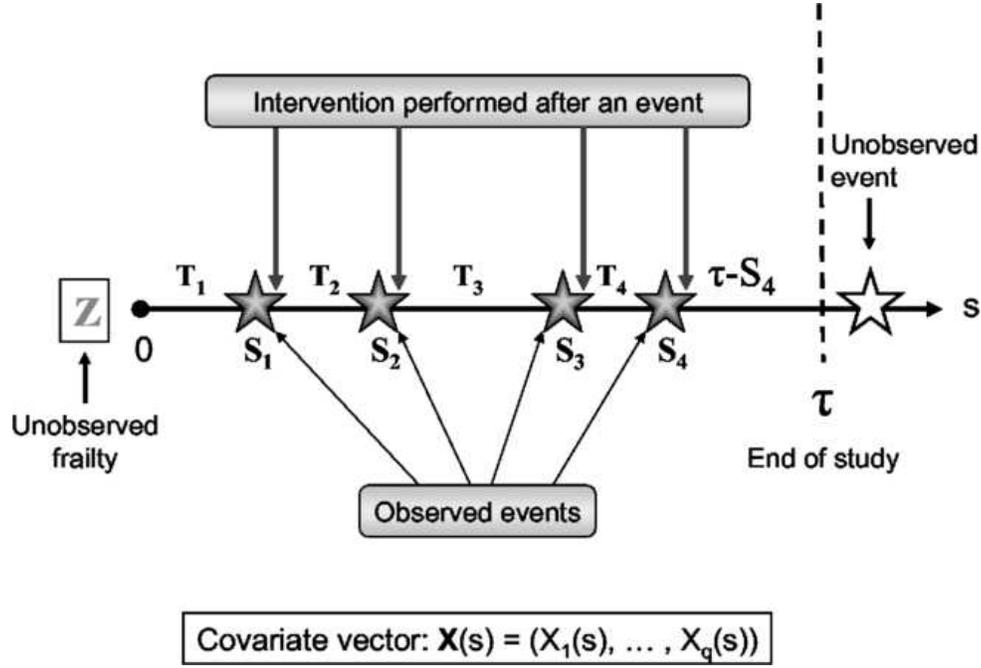

FIG. 1. *Pictorial depiction of the recurrent event data accrual for a unit illustrating the window of observation $[0, \tau]$, intervention performed after an event occurrence, an unobserved frailty $Z$, the presence of a vector of covariates $\mathbf{X}$, the interevent times $T_j$ and the calendar times of event occurrences $S_j$.*

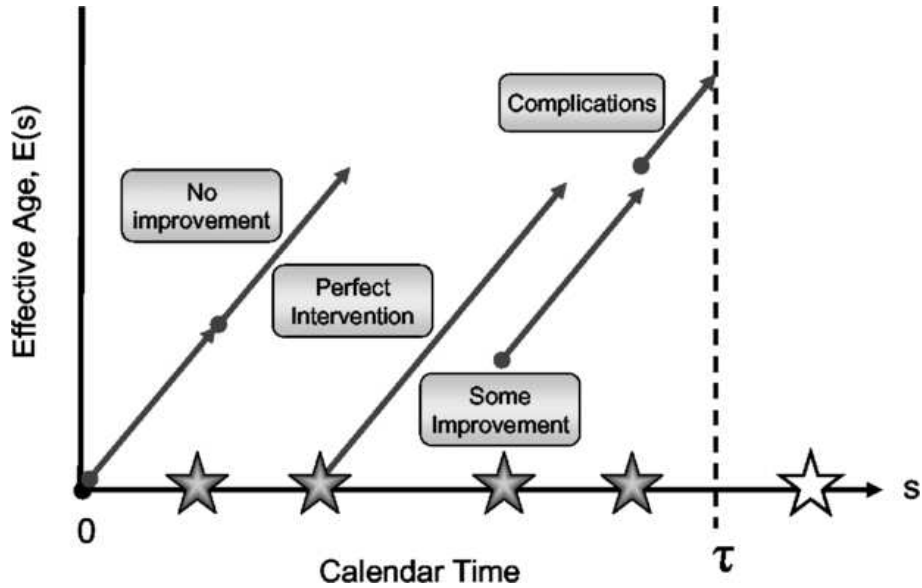

FIG. 2. *An example of an effective age process, $\mathcal{E}(s)$, for a unit encountering successive occurrences of a recurrent event.*

The effective age process could occur in many forms, and the idea is this should be determined in a dynamic fashion in conjunction with interventions that are performed. As such the determination of this process should preferably be through consultations with experts of the subject matter under consideration. Common forms of this effective age process are:

Minimal Intervention or Repair:
$$(3.6) \quad \mathcal{E}(s) = s,$$

Perfect Intervention or Repair:
$$(3.7) \quad \mathcal{E}(s) = s - S_{N^\dagger(s-)},$$

BBS Model:
$$(3.8) \quad \mathcal{E}(s) = s - S_{\Gamma_{\eta(s-)}},$$

where in (3.8) with $I_1, I_2, \ldots$ being independent Bernoulli random variables with $I_k$ having success probability $p(S_k)$ with $p: \Re_+ \to [0,1]$, we define $\eta(s) = \sum_{i=1}^{N^\dagger(s)} I_i$ and $\Gamma_0 = 0$, $\Gamma_k = \min\{j > \Gamma_{k-1}: I_j = 1\}, k = 1, 2, \ldots$. Thus, in (3.8), $\mathcal{E}(s)$ represents the time measured from $s$ since the last perfect repair. This effective age is from Block, Borges and Savits (1985), whereas when $p(s) = p$ for some $p \in [0,1]$, we obtain the effective age process for the Brown and Proschan (1983) minimal repair model. Clearly, the effective age functions (3.6) and (3.7) are special cases of (3.8). Other effective age processes that could be utilized are those associated with the general repair model of Kijima (1989), Stadje and Zuckerman (1991), Baxter, Kijima and Tortorella (1996), Dorado, Hollander and Sethuraman (1997) and Last and Szekli (1998). See also Lindqvist (1999) and Lindqvist, Elvebakk and Heggland (2003) for a review of some of these models pertaining to repairable systems, and González, Peña and Slate (2005) for an effective age process suitable for cancer studies.

A crucial property arising from the intensity specification in (3.5) is it amounts to postulating that, with

$$A^\dagger(s; Z, \lambda_0(\cdot), \alpha, \beta)$$

$$(3.9) \quad = Z \int_0^s Y^\dagger(w) \lambda_0[\mathcal{E}(w)] \cdot \rho[N^\dagger(w-); \alpha] \psi(\mathbf{X}(w)\beta) \, dw,$$

then, conditionally on $Z$, the process

$$(3.10) \quad \begin{aligned} &\{M^\dagger(s; Z, \lambda_0(\cdot), \alpha, \beta) \\ &= N^\dagger(s) - A^\dagger(s; Z, \lambda_0(\cdot), \alpha, \beta): s \geq 0\} \end{aligned}$$

is a square-integrable martingale with PQV process $A^\dagger(\cdot; Z, \lambda_0(\cdot), \alpha, \beta)$. The class of models is general and flexible and subsumes many current models for recurrent events utilized in survival analysis and reliability. In particular, it includes those of Jelinski and Moranda (1972), Gail, Santner and Brown (1980), Gill (1981), Prentice, Williams and Peterson (1981), Lawless (1987), Aalen and Husebye (1991), Wang and Chang (1999), Peña, Strawderman and Hollander (2001) and Kvam and Peña (2005). We demonstrate the class of models via some concrete examples.

EXAMPLE 1. The first example concerns a load-sharing $K$-component parallel system with identical components. The recurrent event of interest is component failure and failed components are not replaced. When a component fails, a redistribution of the system load occurs among the remaining functioning components, and to model this system in a general way, we let $\alpha_0 = 1$ and $\alpha_1, \ldots, \alpha_{K-1}$ be positive constants, referred to as load-share parameters. One possible specification of these parameters is $\alpha_k = K/(K-k), k = 0, 1, 2, \ldots, K-1$, though they could be unknown constants, possibly ordered. The hazard rate of event occurrence at calendar time $s$, provided that the system is still under observation, is $\lambda(s) = \lambda_0(s)[K - N^\dagger(s-)]\alpha_{N^\dagger(s-)}$, where $\lambda_0(\cdot)$ is the common hazard rate function of each component and $N^\dagger(s)$ is the number of component failures observed on or before time $s$. This is a special case of the general class of models with $\mathcal{E}(s) = s$, $\rho(k; \alpha_1, \ldots, \alpha_{K-1}) = [K-k]\alpha_k$, and $\psi(w) = 1$. This is the equal load-sharing model in Kvam and Peña (2005). More generally, this could accommodate environmental or operating condition covariates for the system, and even an unobserved frailty component.

EXAMPLE 2. Assume in a software reliability model that there are $\alpha$ bugs at the beginning of a debugging process and the event of interest is encountering a bug. A possible model is these $\alpha$ bugs are competing to be encountered, and if each of them has hazard rate of $\lambda_0(s)$ of being encountered at time $s$, then the total hazard rate at time $s$ of the software failing is $\lambda_0(s)\alpha$. Upon encountering a bug at time $S_1$, this bug is eliminated, thus decreasing the number of remaining bugs by one. The debugging process is then restarted at time just after $S_1$ (assuming it takes zero time to eliminate the bug, clearly an oversimplification). In general, suppose that just before calendar time $s$, $N^\dagger(s-)$ bugs have been removed, and the last bug was encountered at calendar time $S_{N^\dagger(s-)}$. Then, the overall hazard of encountering a bug at calendar time $s$ with $s > S_{N^\dagger(s-)}$ is $\lambda_0(s - S_{N^\dagger(s-)})[\alpha - N^\dagger(s-)]$. Thus, provided that the debugging process is still in progress at time $s$, then the hazard of encountering a bug at time $s$ is $\lambda(s) = \lambda_0(s - S_{N^\dagger(s-)})\max[0, \alpha - N^\dagger(s-)]$. Again, this is a special case of the general class of models with $\mathcal{E}(s) = s - S_{N^\dagger(s-)}$, a consequence of the restart of the debugging process, $\rho(k; \alpha) = \max\{0, \alpha - k\}$ and $\psi(w) = 1$. This software debugging model is the model of Jelinski and Moranda (1972) and it was also proposed by Gail, Santner and Brown (1980) as a carcinogenesis model. See also Agustin and Peña (1999) for another model in software debugging which is a special case of the general class of models.

Cox's (1972) proportional hazards model is one of the most used models in biomedical and public health settings. Extensions of this model have been used in recurrent event settings, and Therneau and Grambsch (2000) discuss some of these Cox-based models such as the independent increments model of Andersen and Gill (1982), the conditional model of Prentice, Williams and Peterson (1981) and the marginal model of Wei, Lin and Weissfeld (1989). The independent increments model is a special case of the general class of models obtained by taking either $\mathcal{E}(s) = s$ or $\mathcal{E}(s) = s - S_{N^\dagger(s-)}$ with $\rho(k; \alpha) = 1$ and $\psi(w) = \exp(w)$. The marginal model stratifies according to the event number and assumes a Cox-type model for each of these strata, with the $j$th interevent time in the $i$th unit having intensity $Y_{ij}(t)\lambda_{0j}(t)\exp\{X_i(t)\beta_j\}$, where $Y_{ij}(t)$ equals 1 until the occurrence of the $j$th event or when the unit is censored. The conditional model is similar to the marginal model except that $Y_{ij}(t)$ becomes 1 only after the $(j-1)$st event has occurred.

## 4. STATISTICAL INFERENCE

The relevant parameters for the model in (3.5) are $\lambda_0(\cdot)$, $\alpha$, $\beta$ and the parameter associated with the distribution of the frailty variable $Z$. A variety of forms for this frailty distribution is possible, but we restrict to the case where $Z$ has a gamma distribution with mean 1 and variance $1/\xi$. The parameter associated with $G$, the distribution of $\tau$, is usually viewed as a nuisance, though in current joint research with Akim Adekpedjou, a Ph.D. student at the University of South Carolina, the situation where $G$ is informative about the distributions of the interevent times is being explored.

Knowing the values of the model parameters is important because the model can be utilized to predict future occurrences of the event, an important issue especially if an event occurrence is detrimental. To gain knowledge about these parameters, a study is performed to produce sample data which is the basis of inference about the parameters. We consider a study where $n$ independent units are observed and with the observables over (calendar) time $[0, s^*]$ denoted by

$$\text{DATA}_n(s^*)$$
$$(4.11) \quad = \{[(\mathbf{X}_i(v), N_i^\dagger(v), Y_i^\dagger(v), \mathcal{E}_i(v)), v \le s^*],$$
$$i = 1, 2, \ldots, n\}.$$

Observe that $\text{DATA}_n(s^*)$ provides information about the $\tau_i$'s. More generally, we observe the filtrations $\{(\mathcal{F}_{iv}, v \le s^*), i = 1, 2, \ldots, n\}$ or the overall filtration $\mathbf{F}_{s^*} = \{\mathcal{F}_v = \bigvee_{i=1}^n \mathcal{F}_{iv}, v \le s^*\}$. The goals of statistical inference are to obtain point or interval estimates and/or test hypotheses about model parameters, as well as to predict the time of occurrence of a future event, when $\text{DATA}_n(s^*)$ or $\mathbf{F}_{s^*}$ is available. We focus on the estimation problem below, though we note that the prediction problem is of great practical importance.

Conditional on $\mathbf{Z} = (Z_1, Z_2, \ldots, Z_n)$, from (2.3) the likelihood process for $(\lambda_0(\cdot), \alpha, \beta)$ is

$$L_C(s; \mathbf{Z}, \lambda_0(\cdot), \alpha, \beta)$$
$$(4.12) \quad = \prod_{i=1}^n \left\{ Z_i^{N_i^\dagger(s)} \left\{ \prod_{v=0}^s [B_i(v; \lambda_0(\cdot), \alpha, \beta)]^{dN_i^\dagger(v)} \right\} \right.$$
$$\left. \cdot \exp\left[-Z_i \int_0^s B_i(v; \lambda_0(\cdot), \alpha, \beta)\,dv\right] \right\},$$

where $B_i(v; \lambda_0(\cdot), \alpha, \beta) = Y_i^\dagger(v)\lambda_0[\mathcal{E}_i(v)]\rho[N_i^\dagger(v-); \alpha]\psi[\mathbf{X}_i(v)\beta]$. Observe that the likelihood process when the model does not involve any frailties is obtained from (4.12) by setting $Z_i = 1$, $i = 1, 2, \ldots, n$, which is equivalent to letting $\xi \to \infty$. The resulting no-frailty likelihood process is

$$
\begin{aligned}
(4.13) \quad & L(s; \lambda_0(\cdot), \alpha, \beta) \\
&= \prod_{i=1}^n \left\{ \left\{ \prod_{v=0}^s [B_i(v; \lambda_0(\cdot), \alpha, \beta)]^{dN_i^\dagger(v)} \right\} \right. \\
& \quad \left. \cdot \exp\left[ -\int_0^s B_i(v; \lambda_0(\cdot), \alpha, \beta) \, dv \right] \right\}.
\end{aligned}
$$

This likelihood process is the basis of inference about $(\lambda_0(\cdot), \alpha, \beta)$ in the absence of frailties. Going back to (4.12), by marginalizing over **Z** under the gamma frailty assumption, the likelihood process for $(\lambda_0(\cdot), \alpha, \beta, \xi)$ becomes

$$
\begin{aligned}
(4.14) \quad & L(s; \lambda_0(\cdot), \alpha, \beta, \xi) \\
&= \prod_{i=1}^n \left\{ \left( \frac{\xi}{\xi + \int_0^s B_i(w; \lambda_0(\cdot), \alpha, \beta) \, dw} \right)^\xi \right. \\
& \quad \cdot \prod_{v=0}^s \left[ (N_i^\dagger(v-) + \xi) B_i(v; \lambda_0(\cdot), \alpha, \beta) \right. \\
& \quad \left. \left. \cdot \left( \xi + \int_0^s B_i(w; \lambda_0(\cdot), \alpha, \beta) \, dw \right)^{-1} \right]^{dN_i^\dagger(v)} \right\}.
\end{aligned}
$$

There are two possible specifications for the baseline hazard rate function $\lambda_0(\cdot)$: parametric or nonparametric. If parametrically specified, then it is postulated to belong to some parametric family of hazard rate functions, such as the Weibull or gamma family, that depends on an unknown $p \times 1$ parameter vector $\theta$. In this situation, a possible estimator of $(\theta, \alpha, \beta, \xi)$, given $\mathbf{F}_{s^*}$, is the maximum likelihood estimator $(\hat{\theta}, \hat{\alpha}, \hat{\beta}, \hat{\xi})$, the maximizer of the mapping $(\theta, \alpha, \beta, \xi) \mapsto L(s^*; \lambda_0(\cdot; \theta), \alpha, \beta, \xi)$. Stocker (2004) studied the finite- and large-sample properties, and associated computational issues, of this parametric ML estimator in his dissertation research. In particular, following the approach of Nielsen, Gill, Andersen and Sørensen (1992), he implemented an expectation–maximization (EM) algorithm (cf. Dempster, Laird and Rubin, 1977) for obtaining the ML estimate. In this EM implementation the frailty variates $Z_i$ are viewed as missing and a variant of the no-frailty likelihood process in (4.13) is used for the maximization step in this algorithm. We refer to Stocker (2004) for details of this computational implementation. For large $n$, and under certain regularity conditions, it can also be shown that $(\hat{\theta}, \hat{\alpha}, \hat{\beta}, \hat{\xi})$ is approximately multivariate normally distributed with mean $(\theta, \alpha, \beta, \xi)$ and covariance matrix $\frac{1}{n} I^{-1}(\hat{\theta}, \hat{\alpha}, \hat{\beta}, \hat{\xi})$, where $I(\theta, \alpha, \beta, \xi)$ is the observed Fisher information associated with the likelihood function $L(s^*; \lambda_0(\cdot; \theta), \alpha, \beta, \xi)$. That is, with $\Theta = (\theta, \alpha, \beta, \xi)^{\text{t}}$, $I(\theta, \alpha, \beta, \xi) = -\{\partial^2/\partial\Theta \, \partial\Theta^{\text{t}}\} \cdot l(s^*; \lambda_0(\cdot; \theta), \alpha, \beta, \xi)$, where $l(s; \lambda_0(\cdot; \theta), \alpha, \beta, \xi)$ is the log-likelihood process given by

$$
\begin{aligned}
(4.15) \quad & l(s; \lambda_0(\cdot; \theta), \alpha, \beta, \xi) \\
&= \sum_{i=1}^n \left\{ \xi \log\left( \frac{\xi}{\xi + \int_0^s B_i(w; \lambda_0(\cdot; \theta), \alpha, \beta) \, dw} \right) \right. \\
& \quad + \int_0^s \log\left[ (N_i^\dagger(v-) + \xi) \right. \\
& \quad \quad \cdot B_i(v; \lambda_0(\cdot; \theta), \alpha, \beta) \\
& \quad \left. \left. \cdot \left( \xi + \int_0^s B_i(w; \lambda_0(\cdot; \theta), \alpha, \beta) \, dw \right)^{-1} \right] dN_i^\dagger(v) \right\}.
\end{aligned}
$$

Tests of hypotheses and construction of confidence intervals about model parameters can be developed using the asymptotic properties of the ML estimators. For small samples, they can be based on their approximate sampling distributions obtained through computer-intensive methods such as bootstrapping. It is usually the case that a parametric specification of $\lambda_0(\cdot)$ is more suitable in the reliability and engineering situations.

In biomedical and public health settings, it is typical to specify $\lambda_0(\cdot)$ nonparametrically, that is, to simply assume that $\lambda_0(\cdot)$ belongs to the class of hazard rate functions with support $[0, \infty)$. This leads to a semiparametric model, with infinite-dimensional parameter $\lambda_0(\cdot)$ and finite-dimensional parameters $(\alpha, \beta, \xi)$. Inference for this semiparametric model was considered in Peña, Slate and González (2007). In this setting, interest is focused on the finite-dimensional parameters $(\alpha, \beta, \xi)$ and the infinite-dimensional parameter $\Lambda_0(\cdot) = \int_0^\cdot \lambda_0(w) \, dw$ and its survivor function $S_0(\cdot) = \prod_{w=0}^\cdot [1 - \Lambda_0(dw)]$. A difficulty encountered in estimating $\Lambda_0(\cdot)$ is that in the intensity specification in (3.5), the argument of $\lambda_0(\cdot)$ is the effective age $\mathcal{E}(s)$, not $s$, and our interest is in $\Lambda_0(t)$, $t \geq 0$. This poses difficulties, especially in establishing asymptotic properties, because the usual martingale approach as pioneered by Aalen (1978), Gill (1980) and Andersen and Gill (1982) (see also Andersen et al., 1993 and Fleming and Harrington, 1991) does not directly carry through. In the simple i.i.d. renewal setting where $\mathcal{E}(s) = s - S_{N^\dagger(s-)}$, $\rho(k; \alpha) = 1$ and $\psi(w) = 1$, Peña, Strawderman and Hollander (2000, 2001), following ideas of Gill (1981) and Sellke (1988), implemented an approach using time transformations to obtain estimators of $\Lambda_0(\cdot)$ and $S_0(\cdot)$. In an indirect way, with partial use of Lenglart's inequality and Rebolledo's MCLT, they obtained asymptotic properties of these estimators, such as their consistency and their weak convergence to Gaussian processes. This approach in Peña, Strawderman and Hollander (2001) was also utilized in Peña, Slate and González (2007) to obtain the estimators of the model parameters in the more general model.

The idea behind this approach is to define the predictable (with respect to $s$ for fixed $t$) doubly indexed process $C_i(s,t) = I\{\mathcal{E}_i(s) \leq t\}, i = 1, 2, \ldots, n$, which indicates whether at calendar time $s$ the unit's effective age is at most $t$. We then define the processes $N_i(s,t) = \int_0^s C_i(v,t) N_i^\dagger(dv)$, $A_i(s,t) = \int_0^s C_i(v,t) \cdot A_i^\dagger(dv)$, and $M_i(s,t) = N_i(s,t) - A_i(s,t) = \int_0^s C_i(v,t) M_i^\dagger(dv)$. Because for each $t \geq 0$, $C_i(\cdot; t)$ is a predictable and a $\{0,1\}$-valued process, then $M_i(\cdot, t)$ is a square-integrable martingale with PQV $A_i(\cdot, t)$. However, observe that for fixed $s$, $M_i(s, \cdot)$ is not a martingale though it still satisfies $E\{M_i(s,t)\} = 0$ for every $t$. The next step is to have an alternative expression for $A_i(s,t)$ such that $\Lambda_0(\cdot)$ appears with an argument of $t$ instead of $\mathcal{E}_i(v)$. With $\mathcal{E}_{ij-1}(v) = \mathcal{E}_i(v) I\{S_{ij-1} < v \leq S_{ij}\}$ on $I\{Y_i^\dagger(v) > 0\}$, this is achieved as follows:

$$A_i(s,t; \Lambda_0(\cdot), \alpha, \beta)$$
$$= \int_0^s Y_i^\dagger(v) \rho[N_i^\dagger(v-); \alpha]$$
(4.16)
$$\quad \cdot \psi(X_i(v)\beta) I\{\mathcal{E}_i(v) \leq t\} \lambda_0[\mathcal{E}_i(v)] \, dv$$
$$= \sum_{j=1}^{N_i^\dagger(s-)} \int_{S_{ij-1}}^{S_{ij}} Y_i^\dagger(v) \rho[N_i^\dagger(v-); \alpha]$$
$$\quad \cdot \psi(X_i(v)\beta) I\{\mathcal{E}_{ij-1}(v) \leq t\}$$
$$\quad \cdot \lambda_0[\mathcal{E}_{ij-1}(v)] \, dv$$
$$+ \int_{S_{iN_i^\dagger(s-)}}^s Y_i^\dagger(v) \rho[N_i^\dagger(v-); \alpha]$$
$$\quad \cdot \psi(X_i(v)\beta) I\{\mathcal{E}_{iN_i^\dagger(s-)}(v) \leq t\}$$
$$\quad \cdot \lambda_0[\mathcal{E}_{iN_i^\dagger(s-)}(v)] \, dv.$$

Letting

$$\varphi_{ij-1}(v; \alpha, \beta)$$
$$= \frac{\rho[N_i^\dagger(v-); \alpha] \psi(X_i(v)\beta)}{\mathcal{E}'_{ij-1}(v)} I\{S_{ij-1} < v \leq S_{ij}\},$$

and defining the new "at-risk" process

$$Y_i(s,t; \alpha, \beta)$$
$$= \sum_{j=1}^{N_i^\dagger(s-)} I\{t \in (\mathcal{E}_{ij-1}(S_{ij-1}+), \mathcal{E}_{ij-1}(S_{ij})]\}$$
(4.17)
$$\quad \cdot \varphi_{ij-1}[\mathcal{E}_{ij-1}^{-1}(t); \alpha, \beta]$$
$$+ I\{t \in (\mathcal{E}_{iN_i^\dagger(s-)}(S_{iN_i^\dagger(s-)}+),$$
$$\quad \mathcal{E}_{iN_i^\dagger(s-)}(s \wedge \tau_i)]\}$$
$$\quad \cdot \varphi_{iN_i^\dagger(s-)}[\mathcal{E}_{iN_i^\dagger(s-)}^{-1}(t); \alpha, \beta],$$

then, after a variable transformation $w = \mathcal{E}_{ij-1}(v)$ for each summand in (4.16), we obtain an alternative form of $A_i(s,t)$ given by $A_i(s,t; \Lambda_0(\cdot), \alpha, \beta) = \int_0^t Y_i(s,w; \alpha, \beta) \Lambda_0(dw)$. The utility of this alternative form is that $\Lambda_0(\cdot)$ appears with the correct argument for estimating it. If, for the moment, we assume that we know $\alpha$ and $\beta$, by virtue of the fact that $M_i(s,t; \alpha, \beta)$ has zero mean, then using the idea of Aalen (1978), a method-of-moments "estimator" of $\Lambda_0(t)$ is

(4.18)
$$\hat{\Lambda}_0(s^*, t; \alpha, \beta)$$
$$= \int_0^t \frac{I\{S_0(s^*, w; \alpha, \beta) > 0\}}{S_0(s^*, w; \alpha, \beta)} \sum_{i=1}^n N_i(s^*, dw),$$

where $S_0(s,t) = \sum_{i=1}^n Y_i(s,t; \alpha, \beta)$. This "estimator" of $\Lambda_0(\cdot)$ can be plugged into the likelihood function over $[0, s^*]$ to obtain the profile likelihood of $(\alpha, \beta)$, given by

$$L_P(s^*; \alpha, \beta)$$
(4.19)
$$= \prod_{i=1}^n \prod_{j=1}^{N_i^\dagger(s^*)} [\rho(j-1; \alpha) \psi[X_i(S_{ij})\beta]$$
$$\quad \cdot (S_0(s^*, \mathcal{E}_i(S_{ij}); \alpha, \beta))^{-1}]^{dN_i^\dagger(S_{ij})}.$$

This profile likelihood is reminiscent of the partial likelihood of Cox (1972) and Andersen and Gill (1982) for making inference about the finite-dimensional parameters in the Cox proportional hazards model and the multiplicative intensity model. The estimator of $(\alpha, \beta)$, denoted by $(\hat{\alpha}, \hat{\beta})$, is the maximizer of the mapping $(\alpha, \beta) \mapsto L_P(s^*; \alpha, \beta)$. Algorithms and software for computing the estimate $(\hat{\alpha}, \hat{\beta})$ were developed in Peña, Slate and González (2007). The estimator of $\Lambda_0(t)$ is obtained by substituting $(\hat{\alpha}, \hat{\beta})$ for $(\alpha, \beta)$ in $\hat{\Lambda}_0(s^*, t; \alpha, \beta)$ to yield the generalized Aalen–Breslow–Nelson estimator

$$\hat{\Lambda}_0(s^*, t) = \int_0^t \frac{I\{S_0(s^*, w; \hat{\alpha}, \hat{\beta}) > 0\}}{S_0(s^*, w; \hat{\alpha}, \hat{\beta})} \cdot \sum_{i=1}^n N_i(s^*, dw). \quad (4.20)$$

By invoking the product-integral representation of a survivor function, a generalized product-limit estimator of the baseline survivor function $S_0(t)$ is $\hat{S}_0(s^*, t) = \prod_{w=0}^t [1 - \hat{\Lambda}_0(s^*, dw)]$.

Peña, Slate and González (2007) also discussed the estimation of $\Lambda_0(\cdot)$ and the finite-dimensional parameters $(\alpha, \beta, \xi)$ in the presence of gamma-distributed frailties. The ML estimators of these parameters maximize the likelihood $L(s^*; \Lambda_0(\cdot), \alpha, \beta, \xi)$ in (4.14), with the proviso that the maximizing $\Lambda_0(\cdot)$ jumps only at observed values of the $\mathcal{E}_i(S_{ij})$'s. An EM algorithm finds these maximizers and its implementation is described in detail in Peña, Slate and González (2007). We briefly describe the basic ingredients of this algorithm here.

With $\theta = (\Lambda_0(\cdot), \alpha, \beta, \xi)$, in the expectation step of the algorithm, expressions for $E\{Z_i|\theta, \mathbf{F}_{s^*}\}$ and $E\{\log Z_i|\theta, \mathbf{F}_{s^*}\}$, which are easy to obtain under gamma frailties, are needed. For the maximization step, with $E_{\mathbf{Z}|\theta^{(0)}}$ denoting expectation with respect to $\mathbf{Z}$ when the parameter vector $\theta$ equals $\theta^{(0)} = (\Lambda_0^{(0)}(\cdot), \alpha^{(0)}, \beta^{(0)}, \xi^{(0)})$, we require $Q(\theta; \theta^{(0)}, \mathbf{F}_{s^*}) = E_{\mathbf{Z}|\theta^{(0)}}\{\log L_C(s^*; \mathbf{Z}, \theta^{(0)})\}$, where $L_C(s; \mathbf{Z}, \theta)$ is in (4.12). This $Q(\theta; \theta^{(0)}, \mathbf{F}_{s^*})$ is maximized with respect to $\theta = (\Lambda_0(\cdot), \alpha, \beta, \xi)$. This maximization can be performed in two steps: first, maximize with respect to $(\Lambda_0(\cdot), \alpha, \beta)$ using the procedure in the case without frailties except that $S_0(s, t; \alpha, \beta)$ is replaced by $S_0(s, t; \mathbf{Z}, \alpha, \beta) = \sum_{i=1}^n Z_i Y_i(s, t; \alpha, \beta)$; and second, maximize with respect to $\xi$ a gamma log-likelihood with estimated $\log Z_i$ and $Z_i$. To start the iteration process, a seed value for $\Lambda_0(\cdot)$ is needed, which can be the estimate of $\Lambda_0(\cdot)$ with no frailties. Seed values for $(\alpha, \beta, \xi)$ are also required. Through this EM implementation, estimates of $(\Lambda_0(\cdot), \alpha, \beta, \xi)$ are obtained and, through the product-integral representation, an estimate of the baseline survivor function $S_0(\cdot)$.

## 5. ILLUSTRATIVE EXAMPLES

The applicability of these dynamic models still needs further and deeper investigation. We provide in this section illustrative examples to demonstrate their potential applicability.

EXAMPLE 3. The first example deals with a data set in Kumar and Klefsjö (1992) which consists of failure times for hydraulic systems of load-haul-dump (LHD) machines used in mining. The data set has six machines with two machines each classified as old, medium and new. For each machine the successive failure times were observed and the resulting data is depicted in Figure 3. Using an effective age process $\mathcal{E}(s) = s - S_{N^\dagger(s-)}$, this was analyzed in Stocker (2004) (see also Stocker and Peña, 2007) using the general class of models when the baseline hazard function is postulated to be a two-parameter Weibull hazard function $\Lambda_0(t; \theta_1, \theta_2) = (t/\theta_2)^{\theta_1}$, and in Peña, Slate and González (2007) when the baseline hazard function is nonparametrically specified. The age covariate was coded according to the dummy variables $(X_1, X_2)$ taking values $(0, 0)$ for the oldest machines, $(1, 0)$ for the medium-age machines and $(0, 1)$ for the newest machines. The parameter estimates obtained for a nonparametric and a parametric baseline hazard function specification are contained in Table 1, where the estimates for the parametric specification are from Stocker (2004). The estimates of the baseline survivor function $S_0(\cdot)$ under the nonparametric and parametric Weibull specifications are overlaid in Figure 4. From this table of estimates, observe that a frailty component is not needed for both nonparametric and parametric specifications since the estimates of the frailty parameter $\xi$ are very large in both cases. Both estimates of the $\beta_1$ and $\beta_2$ coefficients are negative, indicating a potential improvement in the lengths of the working period of the machines when they are of medium age or newer, though an examination of the standard errors reveals that we cannot conclude that the $\beta$-coefficients are significantly different from zero. On the other hand, the estimates of $\alpha$ for both specifications are significantly greater than 1, indicating the potential weakening effects of accumulating event occurrences. From Figure 4 we also observe that the two-parameter Weibull appears to be a good parametric model for the baseline hazard function as the nonparametric and parametric baseline hazard function estimates are quite close to each other. However, a

TABLE 1
*Parameter estimates for the LHD hydraulic data for a nonparametric and a parametric (Weibull) specification of the baseline hazard function*

| Parameter estimated | With a nonparametric specification of $\Lambda_0(t)$ | With a parametric specification, $\Lambda_0(t) = (t/\theta_2)^{\theta_1}$ |
| --- | --- | --- |
| $\alpha$ | 1.03 | 1.03 (0.01) |
| $\beta_1$ | $-0.09$ | $-0.14$ (0.20) |
| $\beta_2$ | $-0.05$ | $-0.08$ (0.20) |
| $\xi$ | $1.54 \times 10^{63}$ | 164198 (1307812) |
| $\theta_1$ | NA | 0.97 (0.075) |
| $\theta_2$ | NA | 0.006 (0.001) |

The values in parentheses in the third column are the approximate standard errors.

formal procedure for validating this claim still needs to be developed. This is a problem in goodness-of-fit which is currently being pursued.

EXAMPLE 4. The second example is provided by fitting the general class of models to the bladder cancer data in Wei, Lin and Weissfeld (1989). A pictorial depiction of this data set can be found in Peña, Slate and González (2007), where it was analyzed using a nonparametric specification of the baseline hazard function. This data set consists of 85 subjects and provides times to recurrence of bladder cancer. The covariates included in the analysis are $X_1$, the treatment indicator (1 = placebo, 2 = thiotepa); $X_2$, the size (in cm) of the largest initial tumor; and $X_3$, the number of initial tumors. In fitting the general model in (3.5) we used $\rho(k; \alpha) =$ $\alpha^k$. Furthermore, since the data set does not contain information about the effective age, we considered two situations for demonstration purposes: (i) perfect intervention is always performed [$\mathcal{E}_i(s) = s - S_{iN_i^\dagger(s-)}$]; and (ii) minimal intervention is always performed [$\mathcal{E}_i(s) = s$]. The parameter estimates obtained by fitting the model with frailties, and other estimates using procedures discussed in the literature, are presented in Table 2. The estimates of their standard errors (s.e.) are in parentheses, with the s.e.'s under the minimal repair intervention model obtained through a jackknife procedure. The other parameter estimates in the table are those from the marginal method of Wei, Lin and Weissfeld (1989), the conditional method of Prentice, Williams and Peterson (1981) and the Andersen and Gill (1982) method, which were mentioned earlier (cf. Therneau and Grambsch, 2000). Estimates of the survivor functions at the mean covariate values are in Figure 5.

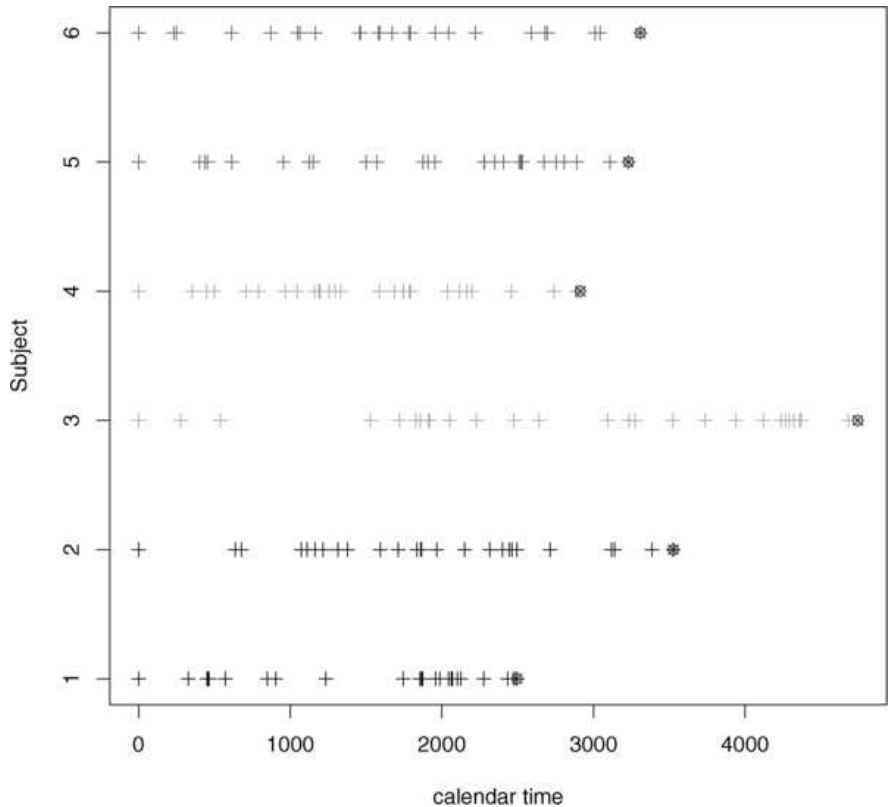

FIG. 3. *Pictorial depiction of the LHD data set from Kumar and Klefsjö (1992) which shows the successive failure occurrences for each of the six machines. Machines 1 and 2 have $(X_1, X_2) = (0, 0)$, machines 3 and 4 have $(X_1, X_2) = (1, 0)$ and machines 5 and 6 have $(X_1, X_2) = (0, 1)$.*

TABLE 2
*Summary of estimates for the bladder data set from the Andersen–Gill (AG), Wei, Lin and Weissfeld (WLW) and Prentice, Williams and Peterson (PWP) methods as reported in Therneau and Grambsch (2000), together with the estimates obtained from the general model using two effective ages corresponding to "perfect" and "minimal" interventions*

| Covariate | Parameter | AG | WLW marginal | PWP conditional | General model Perfect | General model Minimal |
|---|---|---|---|---|---|---|
| $\log N(t-)$ | $\alpha$ | — | — | — | 0.98 (0.07) | 0.79 (0.13) |
| frailty | $\xi$ | — | — | — | $\infty$ | 0.97 |
| rx | $\beta_1$ | $-0.47$ (0.20) | $-0.58$ (0.20) | $-0.33$ (0.21) | $-0.32$ (0.21) | $-0.57$ (0.36) |
| size | $\beta_2$ | $-0.04$ (0.07) | $-0.05$ (0.07) | $-0.01$ (0.07) | $-0.02$ (0.07) | $-0.03$ (0.10) |
| number | $\beta_3$ | 0.18 (0.05) | 0.21 (0.05) | 0.12 (0.05) | 0.14 (0.05) | 0.22 (0.10) |

Observe from this figure that the thiotepa group possesses higher survivor probability estimates compared to the placebo group in either specification of the effective age process. Examining Table 2, note the important role the effective age process provides in reconciling differences among the estimates from the other methods. When the effective age process corresponds to perfect intervention, the resulting estimates from the general model are quite close to those obtained from the Prentice, Williams and Peterson (1981) conditional method; whereas when a minimal intervention effective age is assumed, then the general model estimates are close to those from the marginal method of Wei, Lin and Weissfeld (1989). Thus, the differences among these existing methods are perhaps attributable to the type of effective age process used. This indicates the importance of the effective age in modeling recurrent event data. If possible, it therefore behooves one to monitor and assess this effective process in real applications.

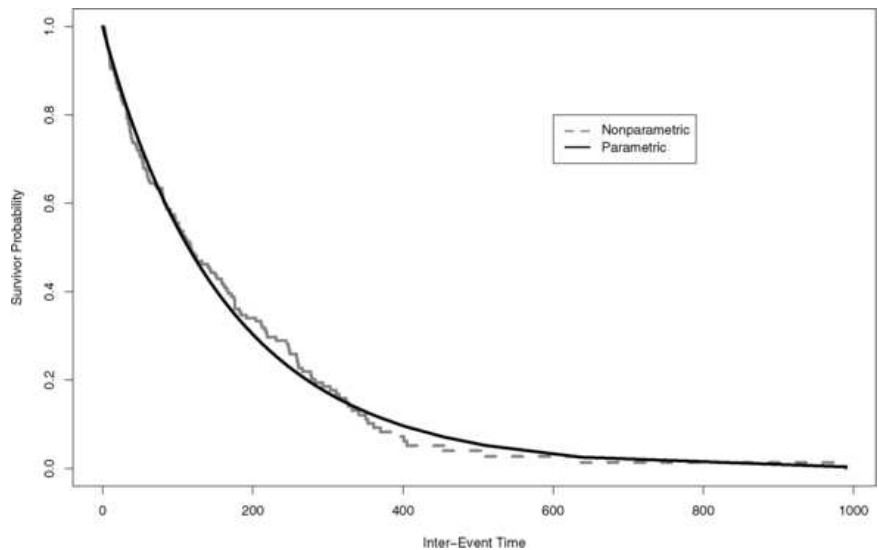

FIG. 4. *Estimates of the baseline survivor function $S_0(t)$ under a nonparametric and a parametric (Weibull) specification for the LHD hydraulic data of Kumar and Klefsjö (1992).*

## 6. OPEN PROBLEMS AND CONCLUDING REMARKS

There are several open research issues pertaining to this general model for recurrent events. First is to ascertain asymptotic properties of the estimators of model parameters under the frailty model when the baseline hazard rate function $\Lambda_0(\cdot)$ is nonparametrically specified. A second problem, which arises after fitting this general class of models, is to validate its appropriateness and to determine the presence of outlying and/or influential observations. This is currently being performed jointly with Jonathan Quiton, a Ph.D. student at the University of South Carolina. Of particular issue is the impact of the sum-quota accrual scheme, leading to the issue of determining the proper sampling distribution for assessing values of test statistics. This validation issue also leads to goodness-of-fit problems. It might, for instance, be of interest to test the hypothesis that the unknown baseline hazard function $\Lambda_0(\cdot)$ belongs to the Weibull class of hazard functions. In current research we are exploring smooth goodness-of-fit tests paralleling those in Peña (1998a, 1998b) and Agustin and Peña (2005) which build on work by Neyman (1937). This will lead to notions of generalized residuals from this general class of models. Another problem is a nonparametric Bayesian approach to failure time modeling. Not much has been done for this approach in this area, though the comprehensive paper of Hjort (1990) provides a solid contribution for the multiplicative intensity model. It is certainly of interest to implement this Bayesian paradigm for the general class of models for recurrent events.

To conclude, this article provides a selective review of recent research developments in the modeling and analysis of recurrent events. A general class of models accounting for important facets in recurrent event modeling was described. Methods of inference for this class of models were also described, and illustrative examples were presented. Some open research problems were also mentioned.

## ACKNOWLEDGMENTS

The author acknowledges the research support provided by NIH Grant 2 R01 GM56182 and NIH COBRE Grant RR17698. He is very grateful to Dr. Sallie Keller-McNulty, Dr. Alyson Wilson and Dr. Christine Anderson-Cook for inviting him to contribute an article to this special issue of *Statistical Science*. He acknowledges his research collaborators, Dr. J. González, Dr. M. Hollander, Dr. P. Kvam,

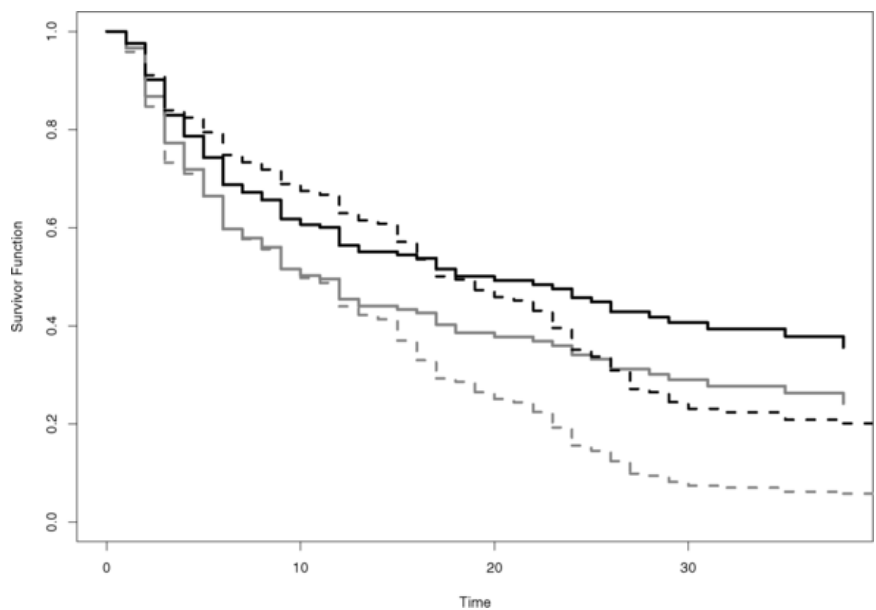

FIG. 5. *Estimates of the survivor functions evaluated at the mean values of the covariates. The solid curves are for the perfect intervention effective age process, whereas the dashed curves are for the minimal intervention effective age process.*

Dr. E. Slate, Dr. R. Stocker and Dr. R. Strawderman, for their contributions in joint research papers on which some portions of this review article are based. He thanks Dr. M. Peña for the improved artwork and Mr. Akim Adekpedjou and Mr. J. Quiton for their careful reading of the manuscript.